\documentclass[preprint, review, 12pt]{article}
\usepackage{setspace}
\usepackage[cm]{fullpage}
\usepackage{color}
\usepackage{array}
\usepackage{lineno}

\usepackage{graphicx}
\usepackage{amssymb}
\usepackage{bm}  % in math mode \bm{} makes {} bold

\begin{document}

\normalsize %\large gives 14 pt

\title{\textbf{Homogeneously saturated model for development in time of the price of an asset}}

%\author{Daniel T. Cassidy}

\author{Daniel T. Cassidy \\ \\ Department of Engineering Physics,  McMaster University, \\ Hamilton, ON, Canada L8S 4L7 \\  \\ cassidy@mcmaster.ca \\}

%\email{cassidy@mcmaster.ca}

 \date{15 July 2011; revised 15 January 2013}% 16 Jan 2013 minor typos revised after submission

\maketitle

\doublespacing
\begin{abstract} % text of abstract
\normalsize

The time development of the price of a financial asset is considered by constructing and solving Langevin equations for a homogeneously saturated model, and for comparison, for a standard model and for a logistic model.  The homogeneously saturated model uses coupled rate equations for the money supply and for the price of the asset, similar to the coupled rate equations for population inversion and power density in a simple model of a homogeneously broadened laser.

\bigskip

Predictions of the models are compared for random numbers drawn from a Student's $t-$distribution.  It is known that daily returns of the DJIA and S\&P 500 indices are fat tailed and are described well by Student's $t-$distributions over the range of observed values. The homogeneously saturated model shows returns that are consistent with daily returns for the indices (in the range of $-30\%$  to $+30\%$) whereas the standard model and the logistic model show returns that are far from consistent with observed daily returns for the indices.  

\bigskip


\end{abstract}

%\keywords{returns; homogeneously saturated; logistic; standard model; Langevin; Student's $t-$distribution; truncation}

%\maketitle % here for {revtex}

\linenumbers

\section{Introduction} %%%%%%%%%%%%%%%%%%%%%%%%%%%%%%%%%% INTRODUCTION %%%%%

The standard model for the development in time of the price of an asset is geometric Brownian motion with a deterministic growth (or drift) rate \cite{Hul, Kel, Bou1}, \cite[Ch 16.4]{Lax}.  The region of support for the normally distributed noise driving term of the Brownian motion is taken as $-\infty$ to $+\infty$, which allows for infinite prices.  This choice of support does not cause difficulty in the standard model as the probability of a large price is essentially zero; the probability of a noise driving term with value $x$ goes as $\exp(-x^2)$.  However, the choice of a `standard' model, which includes an infinite region of support for the noise driving term, makes pricing assets and  pricing options based on these assets difficult when the underlying probability density functions (pdf) is a fat tailed distribution rather than a normal distribution.

It is known that daily returns of the DJIA and the S\&P 500 indices are described by Student's $t-$distributions \cite{Bou1, Cas1} and that Student's $t-$distributions have fat tails.  Hence prices predicted by the `standard' model with a fat tailed distribution for the noise driving term can be very large (essentially infinite) with a non-zero probability.  Here the standard model is defined to include the same region of support of $-\infty$ to $+\infty$ for the noise driving term as for Brownian motion, and the returns are (naively) assumed to follow the distribution that fits the observations over the the full region of support.  For the standard model with an underlying fat tailed distribution, integrals to price European call options diverge \cite{r1, McM, Cas1}.  Thus a standard model with a fat tailed distribution is not adequate, and a standard model with a normal distribution does not match observed daily returns.  

There exist several approaches to price options and assets when the underlying distribution is a fat-tailed distribution.  One approach is to modify the tails of the distribution such that the contributions far into the tails are negligible while not affecting significantly the central portion of the distribution, which fits well the observed data.  This can be accomplished by multiplying the distribution by an $\exp- | \alpha | t^2 $ envelope function \cite{Mor} or by using a generalized $t-$distribution that has terms of $\exp- | \alpha | t^4$ \cite{Lim, Lye}.  It has been demonstrated that an $\exp- | \alpha | t^2 $ envelope multiplication is obtained for a Student's $t-$distribution by truncation of the volatility in the chi-normal mixture that leads to a Student's $t-$distribution \cite{Cas3, Cas4}.

It is possible to price assets and options with a standard model that uses fat tailed distributions by capping the value of the asset or by truncating the underlying pdf.  Using the arbitrage theorem and the constraint that the process be fair (i.e., the price is a martingale), prices for European call options can be found for fat-tailed distributions if the distribution is truncated or if the value of the asset is capped \cite{Cas1, Cas5}.  Truncation or capping keeps integrals, which are needed to price options, finite \cite{r1, McM, Cas1}.  %It is not necessary to solve the Black-Scholes equation to find the price of an option. \cite{Cas1, r1, McM}

Truncation of the pdf for the returns and capping the value of an asset might seem to be {\it ad hoc \/}solutions that allow for any price \cite{r1,McM}.  Truncation is based on conditional probability, which is a sound mathematical concept.  In truncating, one accepts that there is a probability that the value of the asset will exceed the truncation.  The option writer has the ability to select the risk that the value of the underlying asset will exceed the truncation and to price the option for the selected risk.

For capping, both the writer and buyer must agree that the price of the option is based on a maximum value for the asset.  Both sides should recognize that
there is a finite probability that the value of the asset will exceed the agreed upon maximum value but accept that this maximum value will be used.  This controls the
risk, keeps the integrals finite, and allows for the options to be priced.

% 15 Jan 2013 Despite the benefits of capping or of truncating on the prices of assets and of options based on the assets (returns $< -100\%$ or of order $+\infty$ make little physical sense), there is reluctance to accept these approaches.  ARBITRARY Thus it is of interest to look at alternative models for prices of assets and of options. 

A different approach to price assets and options when the underlying distribution is a fat-tailed distribution is to allow for saturation of the price of an asset by depletion of the resource  that supports the price (i.e., by depletion of the reservoir of money that is available to purchase the asset).  This is the approach that is investigated in this paper.  In this paper a homogeneous saturation model for the price of an asset is constructed and compared to the standard model and a logistic model to gain insight into the pricing of financial assets and options based on these assets.  Random numbers drawn from a (fat-tailed) Student's $t-$distribution are used to compare the predictions of the models.

The homogeneous saturation model borrows from laser physics, wherein coupled rate equations are used to describe the interaction between the output (equivalent to price in the pricing of assets considered here) and inversion (equivalent to the reservoir of money available to purchase the asset).   The saturation in a laser keeps the output finite for finite input and ensures that the power output equals the power input.  The saturation in the pricing of an asset keeps the price finite for a finite supply of money, such that the integrals that are required to price an option based on the asset remain finite.

% 15 Jan 2013 The study of pricing options provides some interesting challenges.  In financial markets, noise is significantly greater than the signal, if one regards the drift rate of the price of an asset as the signal \cite{sha}.  This is in contrast to the normal situation in physics, where noise is $<$ signal.  In a brief historical note, Lax \cite[pg 163]{Lax} writes of his start on the study of noise in semiconductor devices.  Lax ``decided that the relevant feature was that the noise consisted in small fluctuations about an operating point.  The operating point was typically a steady state, but since current was drawn it was not an equilibrium state."  In a conductor at room temperature, the drift of an electron under an applied bias is slow compared to the thermal motion.  However, averaging the motion of the electrons over the vast number of electrons ($10^{19}$ cm$^{-3}$) reduces the effect of the thermal motion (noise) to the point that small currents can be measured with precision.  In finance, the drift of the price of a stock is generally obscured by random fluctuations, but there are not $10^{19}$ transactions to average over to obtain precise values for the prices and the distribution of noise might be fat tailed.

A homogeneously saturated model for the development in time of the price of financial assets is investigated in this paper, and is compared to a standard model and a logistic model.  Simple Langevin equations for the time development of the price of an asset, which are first order differential equations with noise driving terms and which should be interpreted as integral equations \cite[pg 172]{Cof} \cite[Ch 10.2]{Lax}, are constructed and solved for the standard model, a logistic model, and for a homogeneously saturated model.  The predictions of the models  are compared for fat tailed noise driving terms.  The logistic model is the standard model with a non-linear saturation term whereas the homogeneous saturation model follows laser physics \cite{Sar, Mil, Cas2}.  In a logistic model, the non-linear saturation term keeps, e.g., the voltage finite in a Van der Pol oscillator \cite[Eq 6, pg 46]{Sar} and populations finite in competitive environments \cite{log, MCM}.  For the homogeneous saturation model, coupled rate equations for the price of an asset and the money supply supporting the asset are postulated and solved.   This coupling between the price and the money supply keeps the price finite and allows for pricing of an European option with fat tailed distributions.  Similarly, the coupling between the inversion and output in a laser keeps the output of the laser finite and equal to the input.  Since only one asset is considered in the simple approach presented in this paper, the money supply is saturated uniformly (homogeneously) by the asset.  It is possible to envision multiple assets interacting with the money supply and with other reservoirs.  In this approach one might allow for inhomogeneous saturation, or homogeneous saturation, or some mixture of the two limiting cases of saturation, and allow for low prices to stimulate purchases, large prices to stimulate sales, and spontaneous decisions.   

%REVIEW OF LITERATURE
%
%\cite{Tot} Bence Toth, Zoltan Eisler, Fabrizio Lillo, Jean-Philippe Bouchaud, Julien Kockelkoren, J. Doyne Farmer How does the market react to your order flow? 4 April 2011	arXiv:1104.0587v1 balance between liquidity takers and liquidity providers
%- market microstructure
%- stability of markets relies on a precise balance between liquidity taking and liquidity providing
%\cite{Far} J.-P. Bouchaud, J. D. Farmer, and F. Lillo.  ÒHow Markets Slowly Digest Changes in Supply and Demand.Ó   \textit{Handbook of Financial Markets: Dynamics and Evolution}, 57-156.  Eds. Thorsten Hens and Klaus Schenk-Hoppe. Elsevier: Academic Press, 2009. 
%- phenomenological approach to building theory
%- study order book, market microstructural; order book is the result of a competition between flow of orders and price dynamics
%- underlines the role of supply and demand;  price is a steady-state not an equilibrium
%
Toth et al. \cite{Tot} and Bouchaud et al. \cite{Far} studied the order book and described the microstructure of the market.  These researchers reported that the stability of markets depends on a precise balance between supply and demand, and that price is a steady state and not an equilibrium.  Smith et al. \cite{Smi} used a rate equation for the density of the order book to understand how prices depend on the rate of flow of orders.  The homogeneously saturated model presented in this paper is consistent with the ideas of these researchers.  The homogeneously saturated model is a phenomenological approach, as is the work of these authors, that uses coupled rate equations to find a balance between supply and price.  The emphasis in this paper is on simple models for the price of a financial asset and not on understanding the microstructure of the market.

Bouchaud and Cont \cite{Bou98} and Bouchaud \cite{Bou99} developed a phenomenological Langevin approach to study market crashes.  They developed a second order differential equation (DE) that is linear in demand minus supply but non-linear in price and solved this DE for the time rate of change of the price of the asset.  The DE for the time rate of change of the price contained a logistic type saturation term.   The derivative of the price went to $-\infty$ once the price of the asset exceeded a threshold value.   It was concluded that crashes are the result of a succession of improbable and unfavourable events, that no precursor to market crashes exists, and that the market behaves as an adaptive system.  The homogeneously saturated model presented in this paper is a phenomenological model and is composed of first order coupled differential equations.  The feedback inherent in the coupled rate equations makes the output of the  homogeneously saturated model adapt to changes in price and money supply.  The emphasis of this paper is on three simple models for the development in time of the price of an asset and not on market crashes.

%\cite{Bou98} J.-P. Bouchaud and R. Cont, A Langevin approach to stock market fluctuations and crashes, Eur. Phys. J. B 6 (1998) 543-550.
%- interested in rare large drops
%- phenomenological Langevin approach; linear in (demand-supply) but non-linear in price; write equation for the time evolution of supply and demand, trend plus random term
%- get a second order DE logistic saturation terms is explained as viscous force
%- derivative of price goes to $-\infty$ in finite time once the price crosses a threshold
%- approximations to solutions (limiting cases for 5 or more variables that describe the market macrostructure), valid for certain ranges of the parameters, to the second order DE 
%- crashes are result of improbable succession of unfavourable events, and have no precursor
%\cite{Bou99} J.-P. Bouchaud, Elements for a theory of financial risk, Physica A 263 (1999) 415--426.
%- Langevin for the time rate of change of the stock price (instantaneous return) with logistic saturation term; explained as risk aversion, which will drop the price to zero once a threshold value is crossed
%- interested crashes
%-market behaves as an adaptive system

Grassia \cite{Gra} added market delay and feedback to obtain a linear second order DE for the price.  Grassia also solved for the time rate of change of the price and thus obtained a Langevin equation.  Grassia studied the time dependence and stability of the price.  He found that quenching ensured long term bounding of the price of the asset.  Richmond and Sabatelli \cite{Ric} developed a Langevin model of interacting agents to understand fluctuations of the prices of financial assets.  They used their results to understand the personal incomes of several countries.  

Anteneodo and Riera \cite{Ant}  developed a non-linear mean reverting Langevin model to study the stochastic dynamics of volatilities.  Anteneodo and Riera showed that additive-mutliplicative processes are required to obtain fat tailed distributions.  In this work the underlying fat tailed distribution is accepted as a fact \cite{Pra, Ger, Cas1, Cas3} and is used with the three models to investigate the pricing of assets.

\bigskip
\section{Standard Model} %%%%%%%%%%%%%%%%%%%%%%%%%%%%%% STANDARD MODEL %%%%%

In the equations that follow, {\it S\/}({\it t\/}) is the value of an asset at time {\it t\/}, {\it S$_{{\rm o}}$\/} = {\it S\/}(0) is the value of the asset at {\it t\/} = 0, $\alpha$ is a drift rate, $\sigma$ is a scale parameter, and {\it f\/}({\it t\/}) is a stochastic process.  

The standard model for the time development of the value of an asset is

\begin{equation}
{\frac{\rm d}{{\rm d} \, t}} S(t)~=~\alpha \, S(t)~+~\sigma \, S(t)\, f\, (t)
\end{equation}
with solution

\begin{equation}
S(t)~=~S_{o} \, \exp \int_{0}^{t} \left ( \alpha ~+~\sigma \, f\, (\eta ) \right ){\rm \,} \, {d}\eta {\rm ~.}
\end{equation}

\indent The return {\it R\/}({\it t\/}) is 

\begin{equation}
\indent R(t)~=~\ln \left ({\frac{S(t)}{S_{o} }} \right )~=~\int_{0}^{t} \left ( \alpha ~+~\sigma \, f\, (\eta ) \right )\,\,{\rm d} \eta {\rm ~.}
\end{equation}

\indent If the integral of the noise wanders to infinity, which is possible with fat-tailed
distributions, then the return goes to infinity.  The equation for the return shows
that the noise contributes to the return and dominates when $\sigma$ {\it f\/}({\it t\/}) $>$ $\alpha$, which
happens routinely since $\alpha$ is typically a small number.

The equation for the time development of the average value for the asset is given by

\begin{equation}
{\frac{\rm d}{{\rm d} \, t}} \overline{S} (t)~=~\left ( \alpha ~+~{\frac{\sigma^{2} }{2}} \right )\, \overline{S} (t)
\end{equation}
with solution

\begin{equation}
\overline{S} (t)~=~S_{o} \, e^{\left ( \alpha ~+~{\frac{\sigma^{2} }{2}} \right )\,\, t}\,.
\end{equation}
The average value tends to infinity as {\it t\/} tends to infinity for $2\,\alpha + \sigma^{{\rm 2}}>0$  . 

The equation for the time development of the average value of the square of the value of the asset is
\begin{equation}
{\frac{\rm d}{{\rm d} \, t}} \overline{S^{2} } (t)~=~2 \left ( \alpha ~+~{\sigma^{2} } \right )\, \overline{S^{2} } (t) \,.
\end{equation}

These equations give the variance of the value of the asset given that the asset was
worth {\it S$_{{\rm o}}$\/} at {\it t\/} = 0 as

\begin{equation}
{\rm Var} (S(t))~=~S_{o}^{2} \,\, e^{(2\, \alpha ~+~\sigma^{2} \, )\, t} ~\left (e^{\sigma^{2} \, t} ~-~1 \right )\,,
\end{equation}
which is the variance for a log-normal distribution.  This is expected, as the
equations for the average values of {\it S\/}({\it t\/}) and {\it S\/}$^{{\rm 2}}$({\it t\/}) are obtained using a Langevin approach with $\langle${\it f\/}({\it t\/})$\rangle$ = 0, $\langle${\it f\/}({\it t\/}$_{{\rm 1}}$){\it f\/}({\it t\/}$_{{\rm 2}}$)$\rangle$ = $\delta$({\it t\/}$_{{\rm 1}}$$-${\it t\/}$_{{\rm 2}}$), and higher order expectations = 0.  The same results for the average values are obtained by  use of Ito's calculus \cite[pg 189]{Lax}.

As {\it t\/} approaches infinity, the variance approaches infinity if $\alpha + \sigma^{{\rm 2}}  > 0$.

It has been assumed that the conditions for the Langevin approach (i.e., that the diffusion coefficients {\it D$_{{k}}$\/} = 0 for {\it k\/} $>$ 2) hold \cite[Ch 10]{Lax} for equations that are driven with noise sources that are distributed as Student's $t-$distributions, or, equivalently, that Ito's calculus holds.  It is known that Student's $t-$distributions fit market returns and that the time correlations for daily returns are approximately delta function correlated.  Thus the assumption is well motivated and seems reasonable in that it is consistent with
observations.

In the following sections the differential equations for the time development of the value of an asset for the logistic and the homogeneous saturation models are solved.  The solutions are then used with Student's {\it t\/}-distributions to investigate prices of assets.  % 15 Jan 2013 It is necessary to truncate or cap fat-tailed distributions to obtain pricing for European call options with the standard model and realistic distributions, i.e., with fat tailed Student's $t-$distributions for daily returns, and it would be of value to explain and to understand the truncation as more than an {\it ad hoc \/}approach.

\bigskip
\section{Logistic Model}  %%%%%%%%%%%%%%%%%%%%%%%%%%%% LOGISTIC %%%%%

The time development of the value for an asset using a logistic model is 

\begin{equation}
{\frac{\rm d}{{\rm d} \, t}} S(t)~=~\alpha \, S(t)~-~\beta S^{2} (t)~+~\sigma \, S(t)\, f\, (t)
\end{equation}
with solution

\begin{equation}
S(t) ~=~{\frac{S_{o} e^{\int_{0}^{t} \alpha \,\, +\,\, \sigma \, f\, (\eta )\,\,{\rm d} \, \eta} }{1~+ ~\beta \, S_{o} \,\, \int_{0}^{t} e^{\int_{0}^{\zeta} \alpha \,\, +\,\, \sigma \, f\, (\eta )\,\,{\rm d} \, \eta} ~~{\rm d} \, \zeta }}\,\,.
\end{equation}
Note that the standard model is obtained from this logistic model in the limit $\beta = 0$.  

The equation for the development in time of the average value of an asset that follows the logistic equation above is

\begin{equation}
{\frac{\rm d}{{\rm d} \, t}} \overline{S} (t)~=~\left ( \alpha ~+~{\frac{\sigma^{2} }{2}} \right )\, \overline{S} (t)~- ~\beta \, {\overline{S} }^{\, 2} (t)
\end{equation}
with solution

\begin{equation}
\overline{S} (t)~=~{\frac{S_{o} \,\, (\alpha ~+~{\frac{\sigma^{2} }{2}} )\,\, e^{\left ( \alpha \,\, +\,\,{\frac{\sigma^{2} }{2}} \right )\,\, t} }{{\, \beta \, S_{o} \,\, \left (e^{\left ( \alpha \,\, +\,\,{\frac{\sigma^{2} }{2}} \right )\,\, t} ~-~1 \right )~+~ \alpha ~+~{\frac{\sigma^{2} }{2}}} }}\,\,.
\end{equation}
The average value of the asset remains finite for all time and approaches the value

\begin{equation}
\lim_{t \rightarrow\infty} ~\overline{S} (t)~=~{\frac{\alpha ~+ ~{\frac{\sigma^{2} }{2}}}{\beta}}
\end{equation}
as $t$ tends to $\infty$. %for $\alpha$ $>$ 0 or $\sigma$$^{{\rm 2}}$ $>$ 2 $\alpha$. 

Figure 1 is comprised of plots of the average value of {\it S\/}({\it t\/}) as a function of time.  The limiting behaviour as {\it t\/} approaches infinity is clear in the figures.  For finite $\alpha$ + $\sigma$$^{{\rm 2}}$/2 the value of {\it S\/}({\it t\/}) saturates to a finite value for large {\it t\/}.  This behaviour is distinct from a random walk where $\int \alpha + \sigma {\it f\/}({\it t\/})$d${\it t\/}$ wanders or drifts to infinity.  The standard model is a random walk and is obtained from this logistic model by setting $\beta=0$.

\begin{figure}[htbp] %%%%% first figure
\centering
\includegraphics[scale = 0.5]{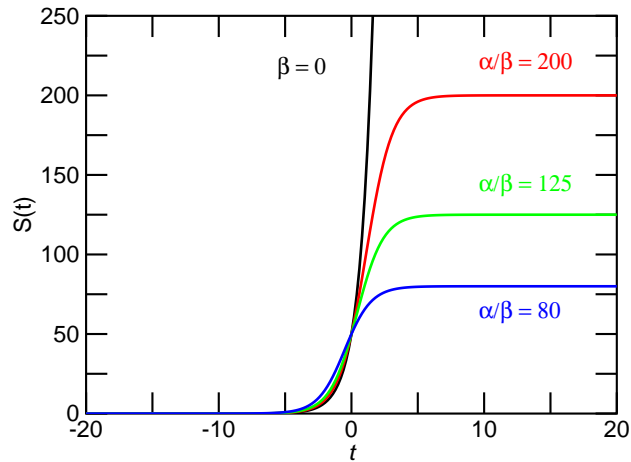}
\caption{{\normalsize Time development of a logistic variable with finite drift for four different values of the saturation parameter.}}
\label{fig:fig1} 
\end{figure} 

The equation for the time development of the square of the value of the asset is 

\begin{equation}
{\frac{\rm d}{{\rm d} \, t}} {\overline{S} }^{\, 2} (t)~=~2 \left ( \alpha ~+~{\sigma^{2} } \right )\, {\overline{S} }^{\, 2} (t)~-~2\, \beta \, {{\overline{S} }^{\, 2} }^{\,{\frac{3}{2}}} (t)
\end{equation}
with solution

\begin{equation}
{\overline{S} }^{\, 2} (t) ~=~{\frac{{S_{o}^{2} }\,\, (\alpha ~+~\sigma^{2} )^{2} }{{\left (e^{-(\alpha \,\, +\,\, \sigma^{2} \,\, )\, t} \,\, (\beta \, S_{o} ~-~\alpha ~-~\sigma^{2} )~-~\beta \, S_{o} \right )^{\, 2} }{\rm }}}~.
\end{equation}

In the limit as {\it t\/} approaches infinity, the variance of {\it S\/}({\it t\/}) under a logistic model remains finite and equals 

\begin{equation}
\lim_{t \rightarrow\infty} {\rm Var} (S(t))~=~{\frac{\alpha ~+~ \sigma^{2} ~+~3 \left ({\frac{\sigma^{2} }{2}} \right )^{2} }{\beta^{\, 2} }} {\rm ~.}
\end{equation}

\indent An approximation for Eq. (9) can be obtained.  The outer integral in the denominator of Eq. (9) presents a challenge, as the value of the integral depends on the values that the stochastic process takes for each point in time.  If one models the stochastic process as small steps in the same direction and of the same magnitude, then the integral can be converted to a summation and this summation can be evaluated.  In this simplified model, which is adopted to allow quick evaluation of the value, the picture then is one of noise as small steps in the same direction, not as abrupt, large jumps.    

Let

\begin{equation}
\int_{0}^{t} \sigma \, f\, (\eta )\,\,{\rm d} \, \eta ~=~W(t)
\end{equation}
and let the Weiner process {\it W\/}({\it t\/}) = {\it W\/}.  Under this simplified picture of the stochastic process,
the integral in the denominator of Eq. (9) can be approximated as

\begin{equation}
\int_{0}^{t} e^{\int_{0}^{\zeta} \alpha \,\, +\,\, \sigma \, f\, (\eta )\,\,{\rm d} \, \eta} ~~{\rm d} \, \zeta ~ \approx{\frac{~t}{N}} \,\, \sum_{i=0}^{N-1} e^{{\frac{\alpha t\,\, +\,\, W }{N}} \, i} ~=~{\frac{t}{N}}{\frac{1~-~e^{\alpha t\,\, +\,\, W} }{1~-~e^{\frac{\alpha t\,\, +\,\, W}{N}} }}\,\,.
\end{equation}

Provided that the interval [0, {\it t\/}] is subdivided into {\it N\/} intervals such that {\it N\/} $>$$>$ $\alpha${\it t\/} +
{\it W\/}, then a series expansion of the exponential in the denominator can be used to
find that 

\begin{equation}
\int_{0}^{t} e^{\int_{0}^{\zeta} \alpha \,\, +\,\, \sigma \, f\, (\eta )\,\,{\rm d} \, \eta} {\rm d} \, \zeta ~ \approx ~{\frac{e^{\alpha t\,\, +\,\, W} ~-~1}{{{\alpha \,\, +\,\, W/t}}}}
\end{equation}
and 

\begin{equation}
S(t)~\approx ~{\frac{S_{o} \, (\alpha \,\, +\,\,{\frac{W}{t}} )\,\, e ^{\alpha t\,\, +\,\, W} }{{\alpha ~+~{\frac{W}{t}} ~+~ \beta \, S_{o} \, \left (e^{\alpha t\,\, +\,\, W} ~-~1 \right )}}}\,\,.
\end{equation}
For large $\alpha${\it t\/} + {\it W\/},

\begin{equation}
\lim_{\alpha t\, +\, W \rightarrow\infty} ~S(t)~=~{\frac{\alpha ~+{\frac{~W}{t}}}{\beta}}
\end{equation}
and the value of the option approaches infinity as $\alpha$ + {\it W\/}/{\it t\/} approaches infinity.  However, the approach is not exponential, as is the approach for the standard
model, Eq. (2).  The return is not linear in $\alpha$ + {\it W/t\/}; the return is the logarithm of $\alpha$ + {\it W/t\/}.  {\it W/t\/} is the average step size (i.e., the total change owing to noise divided by the time taken to make the change).

Figure 2 is a plot of the {\it S\/}({\it x\/}) versus {\it x\/} for {\it x\/} = $\alpha + W(1)$. \ $S(x)$ defined in this manner is
the value of the asset for one day later (i.e., for $t = 1$) with $x$ the value for the
accumulated drift and noise over the one day.  The initial value of the asset, {\it S$_{{\rm o}}$\/}, was
taken as 50.  $\beta$ = 0 gives the standard model.  The three non-zero values of $\beta$ of
0.05/{\it S$_{{\rm o}}$\/}, 0.01/{\it S$_{{\rm o}}$\/}, and 0.02/{\it S$_{{\rm o}}$\/} are the same as in Table 1.  The logistic model
provides some saturation for large values of the drift parameter $\alpha$ and the
accumulated noise $W(1)$ and no saturation for negative values of the accumulated
noise.  It is interesting to contrast the behaviour as a function of $x = \alpha + W(1)$ with the
time development of the solution to a logistic equation.  For the time development
of the solution to a logistic equation, as {\it t\/} approaches infinity, the value of the asset
approaches {\it x\/}/$\beta$.  For a given {\it x\/} that is constant in time, the logistic equation
saturates.  However, if as shown in Fig. 2 {\it x\/} is not limited then the solution to the
logistic equation is not limited.  

\begin{figure}[htbp] %%%%% second figure
\centering
\includegraphics[scale = 0.5]{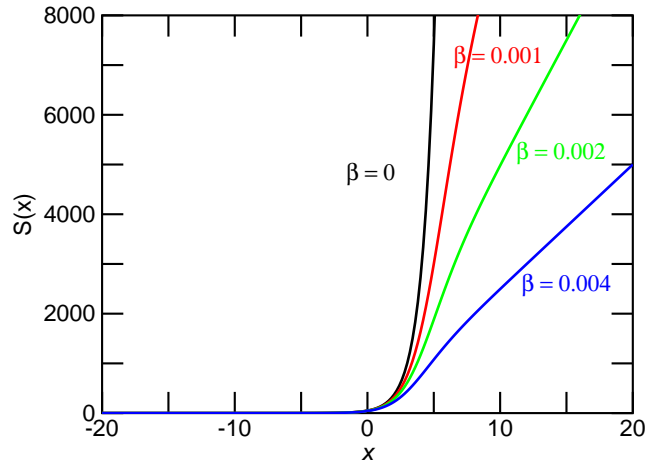}
\caption{{\normalsize Value after one day as a function of the drift plus noise accumulated over one day for different values of the saturation parameter. \ A logistic equation was solved to find the value $S(x)$}}
\label{fig:fig2} 
\end{figure}

Figure 3 shows the return as a function of {\it x \/}where the return is calculated as the
natural logarithm of {\it S\/}({\it x\/})/{\it S\/}$_{{\rm 0}}${\it .  \/}Since {\it S\/}({\it x\/}) was calculated for {\it t\/} =1, the return is the
daily return given the integral of the drift and noise over a time frame of one day. 
The logistic equation shows some saturation of the positive return but no saturation
for a loss.  

\begin{figure}[htbp] %%%%% third figure
\centering
\includegraphics[scale = 0.5]{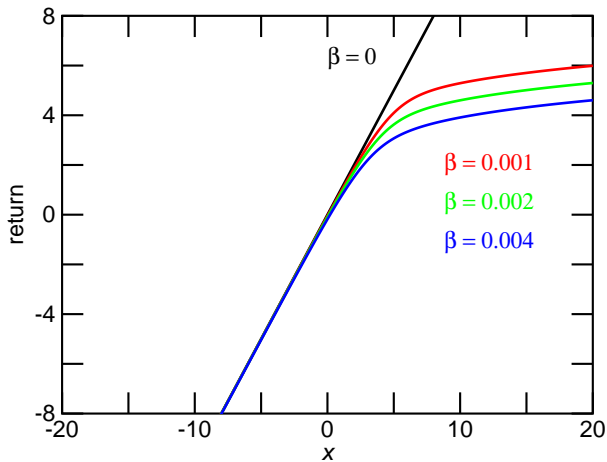}
\caption{{\normalsize Daily return predicted by a logistic equation for various values of the saturation parameter and a function of the drift and the noise accumulated over one day.}}
\label{fig:fig3} 
\end{figure}

Table 1 lists descriptive statistics for simulated values.  Each cell of results contains the ordered pair {\it S\/}(1), {\it R\/}(1) where {\it S\/}(1) is the value of the asset after one day and {\it R\/}(1) is the daily return, defined in Eq. (3) as ln({\it S\/}(1)/{\it S\/}(0)) = ln({\it S\/}(1)/{\it S$_{{\rm o}}$\/}).  To illustrate the effect of the saturation, samples were drawn from a Student's {\it t\/}-distribution with $\nu$ = 2 degrees of freedom multiplied by 10$\times$$\sigma$/sqrt(365) where $\sigma$ is an annualized volatility of 0.3.  Fits to the daily returns for the DJIA and the S\&P 500 show that $\nu$ = 3 would be appropriate \cite{Cas1}.  A drift parameter $\alpha$ = 0.15/365 was assumed.  The initial value for {\it S\/}({\it t\/}) was taken as {\it S$_{{\rm o}}$\/} = 50.

Descriptive statistics for 4096 samples of {\it W\/}({\it t\/}) drawn from a $\nu=2$ Student's {\it t\/}-
distribution and scaled by 10$\times$$\sigma$/sqrt(365) are maximum value = 7.70, minimum
value = $-$6.01, mean = 0.007, standard deviation = 0.43, and kurtosis = 86. 
%\nwln
\begin{center}
\bf{Table 1 Descriptive statistics for simulated values after one day, {\it S\/}(1), and daily
returns, {\it R\/}(1), using Eq. (19) for 4096 samples drawn from a Student's {\it t\/}-distribution with $\nu$ = 2 degrees of freedom, $\alpha$ = 0.0041, $\sigma$ = 0.157, and {\it S$_{{\rm o}}$\/} = 50.0}
\end{center}
%\begin{flushleft}
\begin{tabular}
{|c|c|c|c|c|} \hline
parameter & $\beta$ = 0 & $\beta${\it S$_{{\rm o}}$\/} = 0.05 & $\beta${\it S$_{{\rm o}}$\/} = 0.1 & $\beta${\it S$_{{\rm o}}$\/} = 0.2 \\ \hline
max & 1$\times$10$^{{\rm 5}}$, 7.70 & 7210, 4.97 & 3726, 4.31 & 1895, 3.63 \\ \hline
min & 0.12, $-$6.00 & 0.12, $-$6.01 & 0.12, $-$6.02 & 0.12, $-$6.03 \\ \hline
mean & 104.3, 0.008 & 55.9, $-$0.04 & 51.1,$-$0.092 & 45.4, $-$0.18 \\ \hline
std dev & 2139, 0.43 & 170, 0.39 & 93.4, 0.38 & 50.1, 0.36 \\ \hline
kurtosis & 2237, 86 & 1386, 55 & 1121, 53 & 879, 53 \\ \hline
\end{tabular}
%\begin{flushleft}
\bigskip

The logistic equation shows some limiting of the range of returns for positive returns.  Note that the minimum value of the daily return, {\it R\/}(1), is the same as {\it W\/}({\it t\/}) (cf. the minimum returns for non-zero $\beta$ to the minimum return for $\beta=0$ and to the minimum value for the descriptive statistics for 4096 samples, found in the paragraph preceding Table 1).  The minimum return seems to be unaffected by the logistic equation.  

For the standard model ($\beta$ = 0) the values of an asset for one day later, {\it S\/}(1), lie in
the interval from 0.12 to $1\times10^{{\rm 5}}$ with daily returns, {\it R\/}(1), in the interval of $-600 \%$
to 770 \%.  For a logistic model for the time development of the value of an asset,
{\it S\/}(1) lies in the interval of 0.12 to 3276 with daily returns of $-602 \%$ to 431 \%.  Returns of $< -100 \%$ make little sense.  The
logistic model reduces the maximum value of the asset owing to noise.  The
logistic model also affects the mean value.  Note that the mean reduces from 0.008
to $-0.18$ as $\beta$ is increased from zero.

\section{Homogeneous Saturation} %%%%%%%%%%%%%%%%%%%%%%%%%%  HOMOGENEOUS SATURATION %%%%%

The logistic equation can be justified from a rate equation analysis.

Let {\it M\/}({\it t\/}) be the amount of money that is available to invest in an asset.  Let {\it N\/} be
the rate at which money is pumped into the reservoir of money {\it M\/}({\it t\/})  that can be
used to purchase the asset and let $\beta$$\times${\it M\/}({\it t\/})$\times${\it S\/}({\it t\/}) be the rate that money is removed
from the reservoir owing to purchases of the asset.  Let $\tau$ be a characteristic time
constant that allows for money to be removed or added to the reservoir, depending
on whether {\it M\/}({\it t\/}) is greater than or less than some value {\it M$_{{\rm o}}$\/}.  At this level of
discussion the value of {\it M$_{{\rm o}}$\/} is immaterial as {\it M$_{{\rm o}}$\/}/$\tau$ can be combined with {\it N\/}.  

A rate equation for {\it M\/}({\it t\/}) is then

%\end{flushleft}

\begin{equation}
{\frac{\rm d}{{\rm d} \, t}} M(t)~=~N~-~\beta \times  S(t) \times  M(t)\, ~+~{\frac{M(t)~-~M_{o} }{\tau }}\,\,.
\end{equation}
%\begin{flushleft}

All parameters in the rate equation have a time dependence, but it is assumed that  these parameters change slowly in time owing to inertia in the system and to
gradual evolution of tastes and performance with time.   Thus each point in time is assumed to evolve about a steady state but it is accepted that this steady state point will also evolve
slowly.  In the concepts of Lax \cite[pg 162]{Lax}, fluctuations about an operating point are considered, but the operating point is a point of steady state and not a point of equilibrium.  In steady state, the time derivative equals zero and 
%\end{flushleft}

\begin{equation}
M(t) ~=~{\frac{N ~+{\frac{~M_{o} }{\tau }}}{{\frac{1}{\tau }} ~+~\beta \, S(t)}} ~=~{\frac{\alpha}{1~+~{\frac{\beta}{\alpha}} \, S(t)}}\,\,.
\end{equation}
%\begin{flushleft}
The second form for {\it M\/}({\it t\/}) has been recast so that leading terms in a series
expansion of
%\end{flushleft}

\begin{equation}
M(t) \times S(t) ~=~ \alpha S(t)~-~\beta S^{2} (t) ~+~{\frac{\beta^{2} }{\alpha}} \, S^{3} (t)~-~{\frac{\beta^{3} }{\alpha^{2} }} \, S^{4} (t) ~+ . . .
\end{equation}
\begin{flushleft}
give a logistic equation for the time development of an asset when
\end{flushleft}

\begin{equation}
{\frac{\rm d}{{\rm d} \, t}} S(t)~=~M(t)\, S(t)~+~\sigma \, S(t)\, f\, (t){\rm ~.}
\end{equation}
%\begin{flushleft}
A closed form solution for the equation above for {\it S\/}({\it t\/}) using the full form for {\it M\/}({\it t\/}),
Eq. (22), does not appear to exist.

A noise term is traditionally added to the value of the stock.  However, noise can
be added to the rate equation for the money available to invest in an asset.  This
approach seems fundamental in that it is the amount of money chasing an asset that
determines the value of an asset.  The value of a liquid asset is a visible and easily
obtained attribute but perhaps not a fundamental quantity.

 \bigskip
If one adds a noise term to the rate equation for {\it M\/}({\it t\/}), Eq. (21), then in steady state 
%\end{flushleft}

\begin{equation}
M(t) ~=~{\frac{N ~+{\frac{~M_{o} }{\tau }} ~+~\sigma \, f\, (t)}{{\frac{1}{\tau }} ~+~\beta \, S(t)}} ~=~{\frac{(\alpha ~+~\sigma \, f\, (t))}{1~+~\beta \, S(t)}} {\rm ~.}
\end{equation}
%\begin{flushleft}

In the last form for {\it M\/}({\it t\/}) the symbols have been redefined.  The equation for the
time development of the value of an asset becomes 
%\end{flushleft}

\begin{equation}
{\frac{\rm d}{{\rm d} \, t}} S(t)~=~M(t)\, S(t)~~=~{\frac{\alpha \, S(t) ~+~\sigma \, S(t)\, f\, (t)}{1~+~\beta \, S(t)}} \,\,.
\end{equation}
%\begin{flushleft}

In this approach the noise is ascribed to fluctuations in the amount of money available to invest in the asset.  The value of the asset under this picture is 
%\end{flushleft}

\begin{equation}
S(t) ~=~{\frac{S_{o} \,\,\, e^{\int_{0}^{t} \alpha \,\, +\,\, \sigma \, f\, (\eta )\,\,{\rm d} \, \eta} }{e^{\beta \, (S(t)~- ~S_{o} )} }} ~=~{\frac{S_{o} \,\, e^{\alpha \, t\,\, +\,\, W(t)\,\,} }{e^{\beta \, (S(t)~-~S_{o} )} }} ~{\rm ,}
\end{equation}
%\begin{flushleft}
a transcendental equation that can be solved for {\it S\/}({\it t\/}) given $\alpha$, $\beta$, {\it S$_{{\rm o }}$\/}, and {\it W\/}({\it t\/}).  The
simple dependence on {\it W\/}({\it t\/}) makes simulation straight forward.

Figures 4 and 5 show {\it S\/}({\it x\/}) and {\it R\/}({\it x\/}) for as a function of {\it x\/} = $\alpha${\it t\/} + {\it W\/}({\it t\/}) with {\it t\/} = 1 for four values of $\beta$.  The case $\beta$ = 0 corresponds to the standard model, Eq. (1).  The saturation denominator limits the range of values for the asset {\it S\/}({\it t\/}) and for the return {\it R\/}({\it t\/}).  Note that for the case $\beta$ = 0, {\it S\/}(20)/{\it S\/}$_{{\rm 0}}$ = e$^{{\rm 20}}$ = 5$\times$10$^{{\rm 8}}$ and {\it S\/}($-$20)/{\it S\/}$_{{\rm 0}}$ = e$^{{\rm -}{20}}$ = 2$\times$10$^{{\rm -}{9 }}$, whereas for the case $\beta$ = 1.0, {\it S\/}(20)/{\it S\/}$_{{\rm 0}}$ = e$^{{\rm 0.33}}$ = 1.39 and {\it S\/}($-$20)/{\it S\/}$_{{\rm 0}}$ = e$^{{\rm -}{0.49}}$ = 0.61  Clearly the homogeneous saturation effectively limits the return:  the standard model, $\beta$ = 0, predicts increases of 500 $\times$10$^{{\rm 6}}$ times the initial value of the asset whereas a homogeneous saturation model with $\beta$ = 1 predicts an increase of 1.39 times the initial value of the asset, both for an accumulated drift and noise of $x = 20$.

\begin{figure}[htbp] %%%%% fourth figure
\centering
\includegraphics[scale = 0.5]{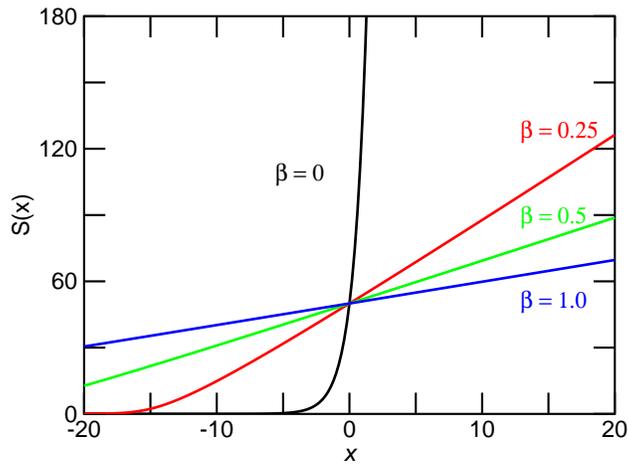}
\caption{{\normalsize Value after one day, $S(x)$, as a function of the drift and noise accumulated over one day, $x$, and for various values of the saturation parameter, $\beta$. \ $\beta = 0$ corresponds to the standard model. \ The value $S(x)$ was determined by solving a homogeneous saturation equation.}}
\label{fig:fig4} 
\end{figure} 

\begin{figure}[htbp] %%%%% fifth figure
\centering
\includegraphics[scale = 0.5]{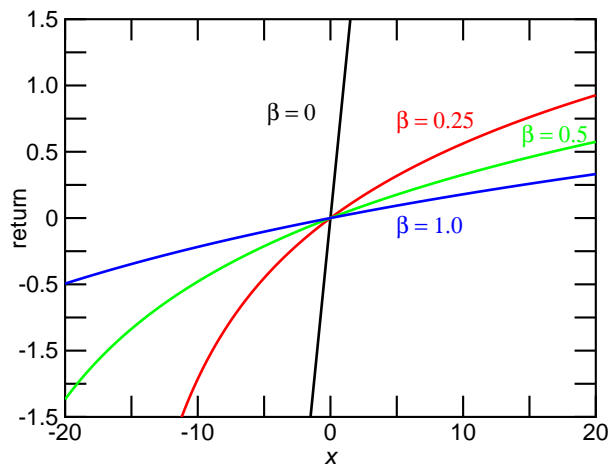}
\caption{{\normalsize Daily return as predicted by a homogeneous saturation equation and as a function of the total drift and noise $x$ accumulated over one day. \ $\beta = 0$ corresponds to the standard model.}}
\label{fig:fig5} 
\end{figure}

The descriptions for {\it M\/}({\it t\/}) and {\it S\/}({\it t\/}) shown in Eqs. (25) and (27) are similar to a
simple description of a homogeneously broadened laser amplifier \cite{Cas2}.  It is possible to
find analytic (albeit transcendental) solutions for the power in each mode in the
case of a simple description of a homogeneously broadened gain medium \cite{Cas2}.  Thus, it
might be possible to solve for multiple assets interacting with the same reservoir of
money for different $\alpha$ and different $\beta$ for each asset.

Table 2 lists descriptive statistics for simulated values.  Each cell of results contains the ordered pair
{\it S\/}(1),  {\it R\/}(1) where {\it S\/}(1) is the value of the asset after one day and {\it R\/}(1) is the daily
return, defined in Eq. (3) as ln({\it S\/}(1)/{\it S\/}(0)) = ln({\it S\/}(1)/{\it S$_{{\rm o}}$\/}).  To illustrate the effect of
the saturation, samples were drawn (as for the results presented in Table 1) from a
Student's {\it t\/}-distribution with $\nu$ = 2 degrees of freedom multiplied by
10$\times$$\sigma$/sqrt(365) where $\sigma$ is an annualized volatility of 0.3.  Fits to the daily returns
from the DJIA and from the S\&P 500 show that $\nu$ = 3 would be appropriate \cite{Cas1}.  A
drift parameter $\alpha$ = 0.15/365 was assumed.  The initial value for {\it S\/}({\it t\/}) was taken as
{\it S$_{{\rm o}}$\/} = 50.

%\end{flushleft}
\begin{center}
\bf{Table 2 Descriptive statistics for simulated values after one day, {\it S\/}(1), and daily returns, {\it R\/}(1), using Eq. (27) for 4096 samples drawn from a Student's {\it t\/}-distribution with $\nu$ = 2 degrees of freedom, $\alpha$ = 0.0041, $\sigma$ = 0.157, and {\it S$_{{\rm o}}$\/} = 50.0}
\end{center}
%\begin{flushleft}
\begin{tabular}
{|c|c|c|c|c|} \hline
parameter & $\beta$ = 0 & $\beta$ = 0.25 & $\beta$ = 0.5 & $\beta$ = 1.0 \\ \hline
max & 1$\times$10$^{{\rm 5}}$, 7.70 & 79.0, 0.46 & 64.9, 0.26 & 57.6, 0.14 \\ \hline
min & 0.12, $-$6.00 & 28.3, $-$0.57 & 38.5, $-$0.26 & 44.1, $-$0.13 \\ \hline
mean & 104.3, 0.008 & 50.0, $-$0.0001 & 50.0,$-$0.0002 & 50.0, $-$0.0001 \\ \hline
std dev & 2139, 0.43 & 1.58, 0.031 & 0.82, 0.016 & 0.42, 0.008 \\ \hline
kurtosis & 2237, 86 & 88, 79 & 86, 76 & 86, 79 \\ \hline
\end{tabular}
\bigskip

The data in the Table 2 clearly shows that the saturation included in Eq (27) limits
the effect of the noise on the value of {\it S\/}({\it t\/}).  With the standard model ($\beta$ = 0) the
maximum value of {\it S\/}(1) is 1$\times$10$^{{\rm 5}}$ with a maximum daily return of 770\% whereas
with $\beta$ = 0.25 the maximum value of {\it S\/}(1) is 79.0 with a maximum daily return of
46\%.  Thus saturation tends to limit the range of daily returns.

It is interesting to note that over the history of the DJIA and the S\&P 500 the daily
returns are in an approximately symmetric interval of $-$0.30 to +0.30 \cite{Cas1}.  The entries
in Table 2 for non-zero $\beta$ show a truncation of the returns in an approximately
symmetric interval.  Given the maximum magnitudes of the returns for the DJIA
and the S\&P 500, it appears that a value of $\beta$ = 0.5 would be appropriate.

It is also interesting to note that the homogeneous saturation does not affect the
mean value, unlike the logistic saturation that does decrease the mean value as the
magnitude of the saturation parameter $\beta$ is increased.

An approximation to the expression for {\it S\/}({\it t\/}) can be made by substituting {\it S\/}({\it t\/}) = {\it S$_{{\rm o}}$\/}
exp($\alpha$t + {\it W\/}({\it t\/})) and making a Taylor series approximation for the exponential term
in the denominator.  The result is
%\end{flushleft}

\begin{equation}
S(t) ~=~{\frac{S_{o} \,\, e^{\alpha \, t\,\, +\,\, W(t)\,\,} }{e^{\beta \, (S(t)~-~S_{o} )} }} ~ \approx  ~{\frac{S_{o} \,\, e^{\alpha \, t\,\, +\,\, W(t)\,\,} }{1~+~ \beta \, S_{o} \,\, e^{\alpha \, t\,\, +\,\, W(t)\,\,} ~-~\beta \, S_{o} }}\,\,.
\end{equation}
%\begin{flushleft}
In this approximation, for large $\alpha${\it t\/} + {\it W\/}({\it t\/}) the value of {\it S\/}({\it t\/}) approaches 1/$\beta$.  The
approximation for {\it S\/}({\it t\/}) above satisfies the differential equation (DE)
%\end{flushleft}

\begin{equation}
{\frac{\rm d}{{\rm d} \, t}} S(t)~=~\left (\alpha ~+~\sigma \times  f\, (t) \right ) \times  S(t) \times  \left (1~-~\beta \times  S(t) \right ){\rm ~.}
\end{equation}
%\begin{flushleft}
In this case the saturation term saturates both the noise and the drift term.  The
form of the saturation suggests the first term of a series expansion of a saturation
denominator.  The solution to this DE has interesting behaviour.  The standard
model is obtained for $\beta$ = 0, whereas for $\beta${\it S$_{{\rm o}}$\/} = 1, {\it S\/}({\it t\/}) = {\it S\/}(0) = {\it S$_{{\rm o}}$\/} for all {\it t\/}, and for
$\beta${\it S$_{{\rm o}}$\/} $>$ 1 and $\alpha${\it t\/} + {\it W\/}({\it t\/}) $<$ 0, negative values of S(t) are obtained.  Thus 0 $\leq$ $\beta${\it S$_{{\rm o}}$\/} $<$ 1.

Table 3 gives descriptive statistics for simulations.  Each cell of results contains the ordered pair {\it S\/}(1),  {\it R\/}(1).  Similar parameters were used in the simulations for Table 3 as were used in the simulations for Tables 1 and 2.  It is interesting to note that the approximate formula, Eq. (28), does not yield a symmetric interval for the daily return.  The data shown in Table 3 show that the minimum return is not greatly affected by the saturation term whereas the maximum return is greatly affected.  Thus whereas Eq. (28) might be simple and computationally efficient, the equation might not be an adequate description of losses. 

%\end{flushleft}
\begin{center}
\bf{Table 3 Descriptive statistics for simulated values after one day, {\it S\/}(1), and daily returns, {\it R\/}(1), using Eq. (28) for 4096 samples drawn from a Student's {\it t\/}-distribution with $\nu$ = 2 degrees of freedom, $\alpha$ = 0.0041, $\sigma$ = 0.157, and {\it S$_{{\rm o}}$\/} = 50.0.}
\end{center}
%\begin{flushleft}
\begin{tabular}
{|c|c|c|c|c|} \hline
parameter & $\beta$ = 0 & $\beta${\it S$_{{\rm o}}$\/} = 0.4 & $\beta${\it S$_{{\rm o}}$\/} = 0.8 & $\beta${\it S$_{{\rm o}}$\/} = 0.9 \\ \hline
max & 1$\times$10$^{{\rm 5}}$, 7.70 & 125, 0.92 & 62.5, 0.22 & 55.6, 0.11 \\ \hline
min & 0.12, $-$6.00 & 0.21, $-$5.49 & 0.61, $-$4.41 & 1.2, $-$3.72 \\ \hline
mean & 104.25, 0.008 & 50.5, $-$0.001 & 49.73,$-$0.011 & 49.77, $-$0.007 \\ \hline
std dev & 2139, 0.43 & 9.6, 0.25 & 3.8, 0.14 & 2.6, 0.10 \\ \hline
kurtosis & 2237, 86 & 125, 117 & 50, 400 & 138, 658 \\ \hline
\end{tabular}

\bigskip
\bigskip
\section{Discussion} %%%%%%%%%%%%%%%%%%%%%%%%%%%% DISCUSSION %%%%%%

The standard model for the development in time of the value of an asset is S({\it t\/}) = {\it A\/}({\it t\/}) $\exp\sigma$$\int$$\xi$({\it t\/})d{\it t\/} where $\xi$({\it t\/}) is
a random variable, $\sigma$ is a positive constant, and {\it A\/}({\it t\/}) includes the drift.  This model predicts an infinite value
for the value of an asset if the accumulated value of the noise $\sigma\int \xi(t)$d$t$ approaches
infinity.  Clearly an infinite value for an asset is not physical \cite{r1,McM}.  The amount of
wealth is limited.  Thus the standard model is valid only for small $\int$$\xi$({\it t\/})d{\it t\/} and the
standard model should not used to predict values of assets for large $\int$$\xi$({\it t\/})d{\it t\/} .

Unfortunately, daily returns are known to have fat tails; Student's {\it t\/}-distributions are known to fit daily returns well \cite[pg 88]{Bou1}, \cite{Cas1, Cas5}.  In contrast to a normal pdf, the probability of a large return is non-zero for a fat tailed pdf such as a Student's $t-$distribution.  

To price an European call with the standard model and assuming that $\xi$ = $\int$$\xi$({\it t\/})d{\it t\/} follows a Student's {\it t\/}-distribution with 3 degrees of freedom, it is necessary to evaluate an integral of the form
%\end{flushleft}

\begin{equation}
\int_{\frac{\ln (K_{T} /A_{T} )}{\sigma}}^{\infty}{\frac{\exp (\sigma \, \xi ) \times\,{\rm d} \xi}{{\left ({1+{\frac{\xi^{2} }{3}}} \right ) ^{\frac{3+1}{2}} } }}
\end{equation}
%\begin{flushleft}
where {\it K$_{T}$\/} is the value of the strike at time {\it T\/} and {\it A$_{T}$\/} is the expected value of the asset at time {\it T\/} \cite{Cas1}.  The integral equals infinity for $\sigma > 0$.  The exponential numerator from the standard model of the price of an asset dominates the power law of the fat tailed distribution.  Thus an infinite value for a European call option is found with fat tailed distributions and the standard model \cite{r1,McM, Cas1, Cas5}.  Figure 6 shows on a logarithmic scale the value of the numerator as dotted lines and the value of the integrand for 0 $\leq$ $\xi$ $\leq$100.  Clearly as $\xi$ tends to infinity, the value of the integrand tends to infinity.

\begin{figure}[htbp] %%%%% sixth figure
\centering
\includegraphics[scale = 0.5]{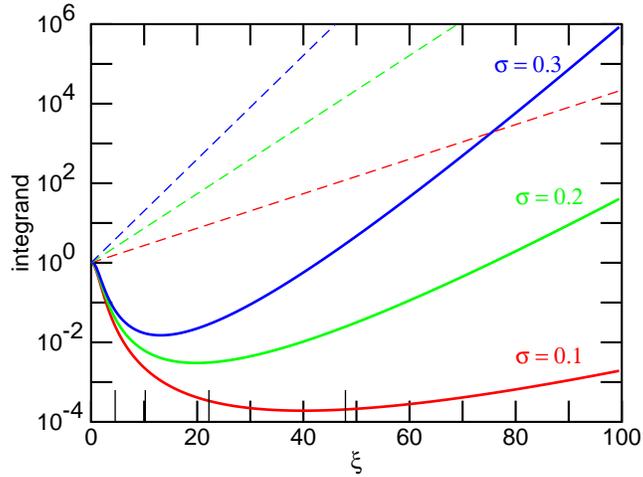}
\caption{{\normalsize Plot of the numerator (broken line) of the integrand and the full integrand (solid line) as a function of the variable of integration for an integral required to price an European call option assuming the return follows a standard model with a fat-tailed noise driving term given by a Student's $t-$distribution with three degrees of freedom and a standard deviation of $\sqrt{3}\sigma$. \ The thin tic marks give the one-tail critical values for probabilities of 0.99, 0.999, 0.9999, and 0.99999 for the $t-$distribution.}}
\label{fig:fig6} 
\end{figure} 

A logistic and a homogeneous saturation model were considered as descriptions for the value of an asset.  Simulations show that the logistic model limits, as compared to the standard model, the positive return whereas the homogeneous saturation model limits both positive and negative returns.   Both positive and negative returns are observed to be limited (i.e., to fall within a range) for the DJIA and for the S\&P 500 indices \cite[see, e.g., Fig.1 ]{Cas1}

The nonlinear saturation provided by the logistic and by particularly the homogeneously saturated model limit the value of an asset such that the integral required to price an asset remains finite.  The value of the asset $S(x)$ appears to increase linearly with $x$ for the homogeneously saturated model as opposed to the exponential increase for $S(x)$ for the standard model (c.f. Fig. 4).  Thus the integral required to price an option, Eq. (30), remains finite for the value of the asset as predicted by a homogeneously saturated model.

A price for an European call option can also be found with the standard model if the value of the asset is capped or if the
distribution is truncated.  The thin tick marks on the abscissa of Fig. 5 give the
critical values {\it x$_{{\rm c}}$ \/}for P\{$\xi$ $\leq$ {\it x$_{{\rm c}}$\/}\} = 0.99, 0.999, 0.9999, and 0.99999 where P\{$\xi$ $\leq$
{\it x$_{{\rm c}}$\/}\}is the probability that $\xi$ $\leq$ {\it x$_{{\rm c }}$\/}.  The critical value for P\{$\xi$ $\leq$ {\it x$_{{\rm c}}$\/}\} = 0.999999 is just
off the right of the chart at $\xi$ = 103.3  Truncation of the pdf at any of the listed critical values will keep the integral in Eq. 30 finite and thus lead to a price for the option with the concomitant level of confidence. 

Truncation of the underlying pdf to obtain a price for an European option has raised some objections.  It is felt that any price can be obtained with truncation and
thus truncation is not a valid approach.

Truncation is a valid approach.  The standard model allows an infinite price, which is not physically possible.  \ Truncation is one method to deal with a mis-specified
model for the value of an asset. \ Another method is to use a model that takes into account a limited reservoir of money that is available to purchase an asset.  The homogeneously saturated model is one such model.  

Truncation also allows to quantify the risk inherent in setting a price for an option.  \ If it is believed that the return on the asset follows a given pdf, then the writer of
the option can select a probability that an asset will exceed a critical value and price based on the selected risk.  This is no different than setting confidence
intervals on experimentally determined numbers.  Once the pdf for the experimentally determined numbers is known, the level of confidence is selected
and this dictates the confidence interval.  If one wishes for 100\% confidence then the confidence interval will be infinite if the region of support is infinite for the
pdf.  If one is willing to accept some risk that the true value might lie outside the confidence interval, then the confidence interval is reduced.  Conversely, one can
select the confidence interval and determine the confidence level that is consistent with the selection.

The logistic and homogeneously saturated models show indirectly that truncation (of the standard model) is a reasonable approach to determine the price of an option.  These alternative models require knowledge of a saturation parameter $\beta$ and the strength of the coupling between the money supply and the price of the asset.  Knowledge of these parameters will be limited.  Thus any option price determined from these alternative models will have uncertainty and a degree of arbitrariness associated with it.  In addition, the noise driving term will need to be be limited or truncated to keep the money supply from approaching infinity.  The uncertainty arising from truncation for the alternative models can be eliminated if the total amount of money available to chase the asset is known.  However, this number will never be known with precision and will be somewhat arbitrary, much like the price of an option as determined from truncation of the pdf in the standard model.  

The homogeneously saturated model that is presented is this paper is not a sophisticated model but it does appear to predict returns that are consistent with returns that are observed.  The model assumes no lower level population, as is appropriate for a simple III-V semiconductor diode laser \cite{Cas2}.  One could include a lower level, i.e, a reservoir of sellers, and transfer population or wealth between buyers and sellers.  Depending on the relative sizes of the reservoirs and the strengths of the couplings between the reservoirs and assets, it might be able to mimmic operation of the markets and thus allow for realistic estimates of prices of assets on the market.  

\bigskip
\section{Conclusion} %%%%%%%%%%%%%%%%%%%%%%%%%%%%%%%%% CONCLUSION %%%%%%

In this paper, a homogeneously saturated model for the time development of the value of an asset has been considered and has been compared to the standard model and a logistic model for the time development of the value of an asset.  These models were based on Langevin equations for the time development of the value of an asset.  The logistic model uses a non-linear (quadratic) saturation term for the price of the asset.  The quadratic saturation term is similar to the saturation term used in population models and in descriptions of Van der Pol's oscillator.  The homogeneously saturation model borrows from lasers physics.  Simple coupled rate equations were constructed and solved in steady state for the time development of the price about the steady state value of the asset and for the time development of the amount of money that is available to purchase the asset.

The standard model is geometric Brownian motion and assumes normal statistics for the noise driving term in the Langevin equation that describes the time development of the price of the asset.  It is known that returns have fat tails and that Student's {\it t\/}-distributions describe well the returns for the DJIA and the S\&P 500 indices.  Hence simulations for the time development of the prices from the three models were performed with samples from a Student's $t-$distribution.

Simulations showed that the logistic model limited the maximum return but not the minimum return.  The homogeneously saturated model limited the magnitudes of both the minimum and maximum returns, and in this respect was, of the three models, the model that was consistent with observed returns for the DJIA and the S\&P 500 indices.

Both the logistic and the homogeneously saturated model require additional parameters, and in particular, a parameter that controls the (non-linear) saturation.  The standard model does not allow for saturation and thus does not require a saturation parameter.  It is likely that this saturation parameter will be somewhat arbitrary and uncertain; it will not likely be known exactly.  Thus this lends some arbitrariness to the alternative models, similar to the arbitrariness that is involved in truncation of the underlying pdf in the standard model to obtain the price of an European call option when using the standard model with fat-tailed distributions.  In this respect, the alternative models indirectly support the approach of pricing options by truncation of the underlying pdf in the standard model.

The homogeneously saturated model, which is presented in this paper, is a simplified, one reservoir model.  The model could be extended to included multiple reservoirs to allow for the interplay between buying, selling, and parking money on the sidelines, and could be extended to include multiple assets interacting with the reservoirs.  These interactions could be homogeneously broadened or inhomogeneously broadened, or a mixture of the limiting forms of the broadening mechanisms.

%\end{flushleft}
\end{document}